\documentclass[journal, twocolumn]{IEEEtran}
\usepackage{tikz,tikz-qtree}	
\usepackage{graphicx}
\usepackage{amsmath,amssymb,bbm}
\usepackage{algorithm,algpseudocode}
\usepackage{url,multirow,hyperref}
\newcommand{\STATE}{\State}

\newcommand{\bld}[1]{{\bf #1}}
\newcommand{\bs}[1]{\boldsymbol{#1}}

\newcommand{\boundellipse}[4]
{(#1) ellipse [x radius = #2, y radius = #3, rotate = #4]}

\newcommand{\openRectangle}[5]
{ 				(#1,#2) 
                -- (#1,#2-#4) 
                -- (#1+#3,#2-#4)node[below,midway] {#5}
                -- (#1+#3,#2)}
\newcommand{\openRectangleDown}[5]
{ 				(#1,#2) 
                -- (#1,#2+#4) 
                -- (#1+#3,#2+#4)node[above,midway] {#5}
                -- (#1+#3,#2)}
                
\newcommand{\vArray}[3]
{ \matrix(m)[matrix of nodes,nodes in empty cells,
  nodes={inner sep=1pt,draw={#1},line width=0.2pt,text width=10pt, text height=10pt,text depth=0.1pt},
  row sep=-0.5\pgflinewidth,
  column sep=-\pgflinewidth,
  ]at (#2,#3){
    $\mu_1$\\
    $\mu_2$\\
     \\
     \\
     \\
     \\
     \\
     \\
   $\mu_K$\\
  };				}   

\usepackage{rotating}
\title{LSTM based AE-DNN constraint for better late reverb suppression in multi-channel LP formulation}
\author{Srikanth~Raj~Chetupalli,
        and Thippur V. Sreenivas\\
        Department of Electrical Communication Engineering, Indian Institute of Science, Bengaluru, 560012.
        }
\begin{document}

\maketitle

\begin{abstract}
 Prediction of late reverberation component using multi-channel linear prediction (MCLP) in short-time Fourier transform (STFT) domain is an effective means to enhance reverberant speech. Traditionally, a speech power spectral density (PSD) weighted prediction error (WPE) minimization approach is used to estimate the prediction filters. The method is sensitive to the estimate of the desired signal PSD. In this paper, we propose a deep neural network (DNN) based non-linear estimate for the desired signal PSD. An auto encoder trained on clean speech STFT coefficients is used as the desired signal prior. We explore two different architectures based on (i) fully-connected (FC) feed-forward, and (ii) recurrent long short-term memory (LSTM) layers. Experiments using real room impulse responses show that the LSTM-DNN based PSD estimate performs better than the traditional methods for late reverb suppression.
\end{abstract}

\section{Introduction}
\label{sec:intro}
Distant speech communication inside an enclosure is adversely affected by the reflection of sound from the walls and the other surfaces (reverberation) \cite{kuttruff2016room}. The strong first few reflection components alter the timbre of the signal, but often aid in improving intelligibility, for example in the absence of a direct component and at longer distances. However, the higher order late reflections distort the spectro-temporal modulations in speech, affecting intelligibility \cite{assmann2004perception, bradley2003importance, gardner1960study}, speech recognition and source localization accuracies \cite{lippmann1997speech, petrick2007harming}. In this paper, we consider suppression of late reflection component of a (static) speech source in a noise free but reverberant environment, using multi-microphone recording.
\par Blind inverse filtering using multi channel linear prediction (MCLP) in the short-time Fourier transform (STFT) domain has been shown to be effective for late reverberation suppression \cite{nakatani2010speech, integrated2009yoshioka}. The reverberant signal is modeled using delayed linear prediction in each frequency bin of STFT, with early reflection component as the desired prediction residual. Maximum likelihood (ML) estimation of the MCLP using a time-varying Gaussian source model has been proposed for parameter estimation. The solution involves iterative and sequential estimation of the desired signal PSD and the prediction coefficients by solving a weighted prediction error (WPE) minimization problem \cite{nakatani2010speech}. However, in the absence of prior knowledge, the sequential ML estimation with reverberant speech based initialization can result in non-monotonic improvement in the desired signal estimation \cite{nakatani2010speech}. Several extensions have been proposed in the literature to improve the performance and the convergence properties of the WPE algorithm. An approach based on smoothed spectral envelope derived using time domain linear prediction is proposed in \cite{introduction2012iwata}. In \cite{multi2015jukic}, a prior estimate based on a complex-generalized Gaussian is proposed to model the heavy-tail distribution of speech STFT. The spectro-temporal nature of speech PSD variation over time is incorporated using a low-rank decomposition approach in \cite{jukic2015multi}. Further, explicit constraint on the variance of late reflection component variance is also explored \cite{constrained2016jukic}. All the above extensions have explored linear estimators and also time-varying nature of speech PSD.
\par In this paper, we explore a non-linear estimation using deep neural network (DNN) for the desired signal PSD. An auto-encoder (AE) trained on clean speech log-magnitude STFT coefficients to give a smoothed PSD at the output, is used as the estimator of the desired signal PSD. This approach is different from the traditional DNN approaches, where the network is trained to predict the clean speech magnitude STFT coefficients \cite{han2015Learning} or a ratio mask \cite{Williamson2017Time} from the reverberant signal STFT coefficients. Our proposed method also differs from the online WPE method using a DNN spectrum estimation proposed in \cite{Kinoshita2017Neural}, where in a DNN is trained to predict directly the PSD of the early component from the reverberant signal STFT. Instead, we use a DNN in tandem with the traditional WPE method to estimate the MCLP filter and hence the residue PSD. The traditional DNN approaches have limited generalizability to un-seen acoustic environments and the source microphone placements. Since we use a clean speech auto-encoder to estimate the speech PSD, the limitation of generalization to unseen acoustic conditions doesn't arise. We explore two DNN architectures for the AE, using (i) fully connected (FC), and (ii) LSTM layers. The experimental results show that, MCLP followed by DNN PSD estimation performs better then earlier methods and also the LSTM architecture performs better than FC architecture.
\section{Multi-channel linear prediction}
\label{sec:ProblemFormulation}
Consider an $M$-channel recording setup of a source signal $s(t)$ inside a reverberant enclosure. The signal $x_m(t)$ recorded at the $m^{th}$ microphone is
\begin{equation}
	x_m(t)=h_m(t) \circledast s(t),
\end{equation}
where $h_m(t)$ relates the acoustic path between the source and the $m^{th}$ microphone position, and consists of two components due to (i) early reflections including the direct path, and (ii) late reflections. In this paper, we consider the suppression of late reflection component at a reference microphone position (ex: the first microphone) given the $M$-channel recordings of a single static source in a interference-free acoustic enclosure. 
\par Let $x_m[n,k]$ denote the short-time Fourier transform (STFT) representation of the microphone signal, where $n,k$ denote the time and frequency bin indices respectively. The signal at the reference microphone ($r=1$) is modeled as,
\begin{equation}\label{eqn:mclpEqn1}
	x_r[n,k]=\sum\limits_{m=1}^{M} \sum\limits_{l=0}^{L-1} g_{m}^*[l,k] x_m[n-D-l,k]+d[n,k],
\end{equation}
where the first term on the right hand side is due to the late reflection component, and $d[n,k]$ is the desired early reflection component at the reference microphone. The delay parameter $D$ controls the chosen boundary between the early and late reflection components of the room impulse response (RIR). In vector form, we can write,
\begin{equation}\label{eqn:mclpEqn3}
	x_r[n,k]=\bld{g}^H[k] \bs{\phi}[n-D,k]+d[n,k],
\end{equation}
where $\bld{g}_{m}[k]=\left[ g_{m}[0,k] \dots g_{m}[L-1,k] \right]^T$, $\bld{g}[k]=\left[ \bld{g}_{1}^T[k] \dots \bld{g}_{M}^T[k] \right]^T$ is the vector of prediction coefficients, and $\bs{\phi}_m[n,k]=\left[x_m[n,k] \dots x_m[n-L+1,k]\right]^T$, $\bs{\phi}[n,k]=\left[ \bs{\phi}_1[n,k]^T \dots \bs{\phi}_M[n,k]^T \right]^T$ is the stacked vector of predictor STFT samples of all the microphones. Given the STFT of $N$ frames of all the mic signals $\{x_m[n,k], 0\leq n \leq N-1, \forall m,k\}$, the goal is to estimate the desired early component signal $\{d[n,k],\forall~n,k\}$ at the reference microphone $r=1$. The MCLP filter $\bld{g}[k]$ is estimated first using a model for the desired signal, and then the early component signal is obtained as the residual of MCLP: $\hat{d}[n,k]=x_1[n,k]-\hat{\bld{g}}^H[k] \bs{\phi}[n-D,k]$.
\par A time-varying Gaussian source model (TVGSM) is proposed in \cite{integrated2009yoshioka}, \cite{nakatani2010speech} for the STFT coefficients of the desired signal,
\begin{equation}\label{eqn:earlyModel}
	d[n,k] \sim \mathcal{N}_c\left( 0,\gamma_{nk} \right),
\end{equation}
where $\gamma_{nk}$ represents the time varying variance of speech due to the changing acoustic, phonetic and prosodic content of speech. However, the STFT coefficients across time and frequency can be assumed independent for a first approximation. Maximum likelihood criterion is used for parameter estimation of the MCLP. From eqns. \eqref{eqn:mclpEqn3}, \eqref{eqn:earlyModel}, the negative log-likelihood $\mathcal{L}(\bld{g},\bs{\gamma})$ can be written as,
\begin{multline}\label{eqn:loglikelihood}
	\mathcal{L}(\bld{g},\bs{\gamma})=\sum\limits_{k=0}^{K/2}\sum\limits_{n=0}^{N-1} \log \gamma_{nk}\\ +\sum\limits_{k=0}^{K/2}\sum\limits_{n=0}^{N-1} (1/\gamma_{nk}) \left|x_1[n,k]-\bld{g}[k]^H \bs{\phi}[n-D,k]\right|^2.
\end{multline}
The parameters $\{\bld{g}[k]\}$ and $\bs{\gamma}$ are estimated alternatively by minimizing $\mathcal{L}(\bld{g},\bs{\gamma})$ iteratively. Minimization of $\mathcal{L}(\bld{g},\bs{\gamma}^{(i)})$ results in a weighted prediction error minimization problem for each $k$, whose solution can be obtained as 
\begin{equation}
	\bld{g}^{(i+1)}[k]=\bld{R}_{\phi\phi}^{-1}[k] \bld{r}_{x\phi}[k],
	\label{eqn:filterEstimate}
\end{equation}
where,
\begin{eqnarray}	
	\bld{R}_{\phi\phi}[k]=\sum\limits_{n=0}^{N-1} \left(1/\gamma_{nk}^{(i)}\right) \bs{\phi}[n-D,k]\bs{\phi}^H[n-D,k],~~\mbox{and} \\\nonumber
	\bld{r}_{x\phi}[k]=\sum\limits_{n=0}^{N-1} \left(1/\gamma_{nk}^{(i)} \right) x_1^*[n,k]\bs{\phi}[n-D,k].
\end{eqnarray}
Similarly the estimate of $\bs{\gamma}$ is obtained by minimizing $\mathcal{L}(\bld{g}^{(i+1)},\bs{\gamma})$, whose solution is,
\begin{equation}
	\gamma_{nk}^{(i+1)}=\left|x_1[n,k]-{\bld{g}^{(i+1)}}^H[k] \bs{\phi}[n-D,k] \right|^{2},~\forall~n,k.
\end{equation}
The initial value $\gamma_{nk}^{(0)}=|x_1[n,k]|^2$ is chosen based on the reverberant signal itself. With no prior knowledge about the speech signal statistics, this choice has been shown to lead to un-realistic estimates of $\gamma_{nk}$, resulting in a degraded residue signal estimate \cite{integrated2009yoshioka}. Hence, we are proposing in this paper, a non-linear estimation of $\gamma_{nk}$ given the MCLP residual signal $\hat{d}[n,k]$, i.e., $\{\hat{\gamma}_{nk}\}=f(\{\hat{d}[n,k]\})$ using an auto-encoder a-priori learned from clean speech log magnitude STFTs.
\section{DNN Estimate of PSD}
\label{sec:proposedMethod}
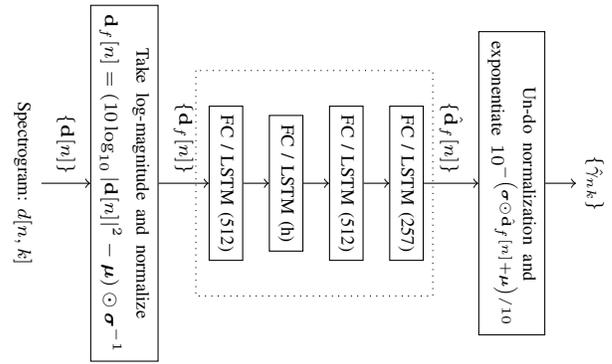
\begin{figure}[h]
\centering
\begin{turn}{270}
\begin{tikzpicture}
	\node[] at (0,-0.1) (in) {\scriptsize Spectrogram: $d[n,k]$};	
	\node[draw,text width=4.5cm,align=center] at (0,1.25) (lm) {\scriptsize Take log-magnitude and normalize $\bld{d}_f[n]=(10\log_{10} |\bld{d}[n]|^2-\bs{\mu})\odot \bs{\sigma}^{-1}$};

	\node[draw,text width=3.8cm,align=center] at (0,6.4) (renorm) {\scriptsize Un-do normalization and exponentiate $10^{-\left(\bs{\sigma} \odot \hat{\bld{d}}_{f}[n]+\bs{\mu}\right)/10}$};
	\node[] at (0,7.5) (out2){\scriptsize $\{\hat{{\gamma}}_{nk}\}$}; 

	\draw[->] (in) --node[left]{\scriptsize $\{\bld{d}[n]\}$} (lm);    
	\draw[->] (renorm) -- (out2);		
	
	\node[draw] at (0,2.6) (ifc) {\scriptsize FC / LSTM  (512)};
	\node[draw] at (0,3.4) (h) {\scriptsize FC / LSTM (h)};
	\node[draw] at (0,4.2) (ofc) {\scriptsize FC / LSTM (512)};
	\node[draw] at (0,5.0) (ae) {\scriptsize FC / LSTM (257)};		

	\draw[dotted] (-1.5,2.2) -- (1.5,2.2) -- (1.5,5.35) -- (-1.5,5.35) -- (-1.5,2.2);

	\draw[->] (lm) --node[left]{\scriptsize $\{\bld{d}_f[n]\}$} (ifc);
	\draw[->] (ifc) -- (h);
	\draw[->] (h) -- (ofc);
	\draw[->] (ofc) -- (ae);	
	\draw[->] (ae) -- node[left]{\scriptsize $\{\hat{\bld{d}}_f[n]\}$} (renorm);		
\end{tikzpicture}
\end{turn}
\caption{DNN auto-encoder for PSD vector non-linear estimation and variance estimation of desired residual signal.}\vspace{-5pt}
\label{fig:blockDiagram}
\end{figure}
We propose a neural network based prior constraint for the estimated variance of the desired signal STFT coefficients. Fig. \ref{fig:blockDiagram} shows a block diagram of the proposed method to estimate $\{\hat{\gamma}_{nk}\}$ within the iterations of the MCLP algorithm. The estimate ${d}[n,k]$, of the early reflection component, obtained using the MCLP filter computed at iteration $i$ is used as the input to DNN. ${d}[n,k]$ is then converted to log-magnitude spectral representation and normalized, before being input to the auto-encoder neural network. The neural network is trained to predict the approximate log-magnitude spectrogram, which is then converted to the power spectrum domain through the inverse normalization, to get a smoothed estimate $\{\hat{\gamma}_{nk}\}$.
\par The auto-encoder considered here can be interpreted as a parameterized function $f(.|\bs{\Phi})$ (defined by the network) trained for faithful reconstruction of the input vector $\bld{d}$  at the output of the network $f(\bld{d}|\bs{\Phi})$, i.e.,
\begin{equation}
	f^*(.|\bs{\Phi})=\underset{f(.|\bs{\Phi})}{\arg\min}~|\bld{d}-f(\bld{d}|\bs{\Phi})|_2^2
\end{equation}
In this paper, we consider two architectures for the network function, (i) fully connected (FC) feed-forward network, and (ii) LSTM network. Both networks comprise of three hidden layers apart from the input and output layers, as shown in Fig. \ref{fig:blockDiagram}. Linear activation is used at the output layer for both FC and LSTM network architectures. For all the other layers, we experimented with different activation functions. For the LSTM network, we experimented with only the activation function of the output gate and not the forget gates. As we show in the experimental section, an exponential linear unit (eLU) activation function is found to give better auto-encoder performance. The number of hidden units is fixed as $512$ for the first and the third hidden layers, and we experiment with different number of units $(h)$ in the bottle-neck layer (second hidden layer). For the FC architecture, we consider input frame expansion with a context of $\pm 2$, i.e., the current frame and two previous and two future frames are used as the input. No such input context is provided for the LSTM architecture, since the network encodes context through the memory states of hidden units. The two networks FC and LSTM are a-priori trained in the same manner. Mean squared error at the output is used as the criterion for optimization of network parameters. AdaDelta optimizer \cite{zeiler2012adadelta} is used for the optimization using the initial learning rate of $0.01$ and number of training epochs is $100$. Keras deep learning framework \cite{chollet2015keras} is used to implement the auto-encoder network.
\begin{algorithm}
	\caption{\emph{Multi channel linear prediction method}}
	\label{alg:mclpAlgo}
	\begin{algorithmic}[1]
		\STATE Input: $\{x_m[n,k],~0\leq n \leq N-1,\forall m\in [1,M],k=[0,K]\}$, $D=2$, $L$, $i_{max}$.
		\STATE Initial estimate $\hat{d}[n,k]=x_1[n,k]$.
		\For {Iteration $i \in [0,i_{max}]$ }
			\STATE Predict $\hat{\bs{\gamma}}=f(\bld{\hat{d}}|\bs{\Phi})$ using the DNN trained a-priori
			\For {$k \in [0,k]$}
				\STATE Estimate the MCLP filter ${\bld{g}^{(i)}}$ using \eqref{eqn:filterEstimate}.
				\STATE Compute the residual 
				\begin{equation*}
					\hat{d}[n,k]=x_1[n,k]-{\bld{g}^{(i)}}^H[k] \bs{\phi}[n-D,k].
				\end{equation*}
			\EndFor
		\EndFor
		\STATE Output $\hat{d}[n,k]$.
	\end{algorithmic}
\end{algorithm}
\par The MCLP algorithm with the proposed DNN PSD constraint is presented in Alg. \ref{alg:mclpAlgo}. The reference microphone signal is taken as the initialization for the first iteration and then estimates for $\gamma_{nk}$ are computed using the pre-trained AE (step-4). The estimated $\{\gamma_{nk}\}$ are used as weights to estimate the prediction filters for each frequency bin $k$, which are then used to compute the residual signal $\hat{d}[n,k]$, used in the next iteration. This estimate is then used to compute the desired signal PSD through the DNN and the procedure is repeated for a pre-fixed number of iterations. 
\section{Experiments and Results}
\label{sec:experimentsAndResults}
 Clean speech sentences from `dr1' set of the TIMIT database are used for training the AE. The dataset consists of speech sentences from $38$ speakers, each speaking $10$ sentences; $7$ sentences from each speaker are used for training, $1$ sentence each for validation and $2$ sentences each for testing. The total number of training sentences is $266$, each of length about $3~sec$. Since, the test set is also drawn from the same set of training speakers, to verify the generalizability of the trained DNN, we also tested using sentences from `dr2' set of the TIMIT database, which contains a total of $760$ utterances ($10$ each from $76$ speakers).
\par RIRs from the REVERB2014 challenge \cite{Kinoshita2016Summary} dataset are used to generate the reverberant signals from the clean speech. The dataset consists of RIRs collected using an $8$ channel uniform circular array (UCA), in three different rooms (RT60=$\{0.25s, 0.6s, 0.73s\}$), at two different distances (near=0.5~m, far=2.0~m) and at two different angles (A=+45, B=-45) with respect to a reference microphone. The STFT analysis is carried out using $32~ms$ window and $75\%$ successive overlap and the delay parameter $D$ is chosen as $2$ frames. We consider a four microphone (alternate microphones in the UCA) sub-array, RIRs from \{room=2, distance='far', angle='A'\} condition and the MCLP order $L=16$ for the all the experiments, unless otherwise stated. Maximum number of iterations of MCLP is chosen to be $5$.
\par We study the performance of the auto-encoder using the average log-spectral difference measure defined as,
\begin{equation}
	LSD(n)=\frac{1}{K} \sum\limits_{k=0}^{K-1} \left| 10\log_{10} \frac{|d[n,k]|^2}{\hat{\gamma}_{nk}} \right|,
\end{equation}
where $|d[n,k]|$ and $|\hat{\gamma}_{nk}|$ denote the magnitude STFT representations at the input and output of the auto-encoder network. Late reverberation suppression performance is measured using average frequency weighted SNR (FwSNR), cepstral distortion (CD), log-likelihood ratio (LLR), signal-to-reverberation-modulation ratio (SRMR) \cite{falk2010NonIntrusice}, and the perceptual measures of PESQ \cite{Rix2001Perceptual} and short-time objective intelligibility (STOI) \cite{taal2010short, SoendergaardMajdak2013}. We compare the performance of proposed approach with the WPE \cite{nakatani2010speech}, CGG \cite{multi2015jukic} methods and also using a time-domain auto regressive (AR) model based smooth PSD estimation (prediction order $21$) \cite{introduction2012iwata}. Speech examples with spectrogram illustrations are available online\footnote{\href{www.ece.iisc.ernet.in/~sraj/lstmMCLP.html}{www.ece.iisc.ernet.in/~sraj/lstmMCLP.html}}.
\begin{figure}[h]
	\centering
	\begin{minipage}[b]{0.48\linewidth}
		\centerline{\includegraphics[width=1.5in,height=1.1in]{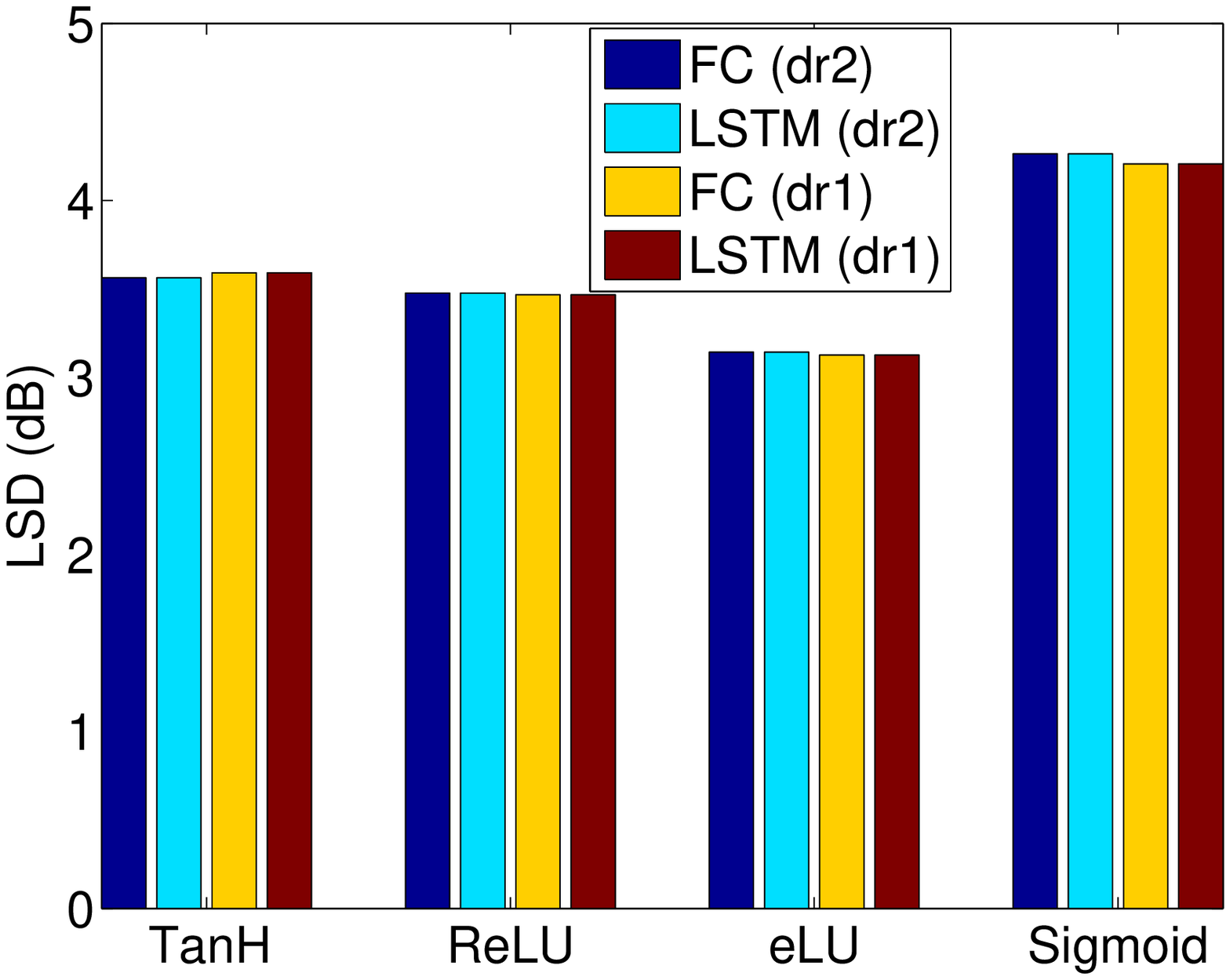}}
		\vspace{-2pt}				
		\centerline{(a)}
	\end{minipage}
	\begin{minipage}[b]{0.48\linewidth}
		\centerline{\includegraphics[width=1.5in,height=1.1in]{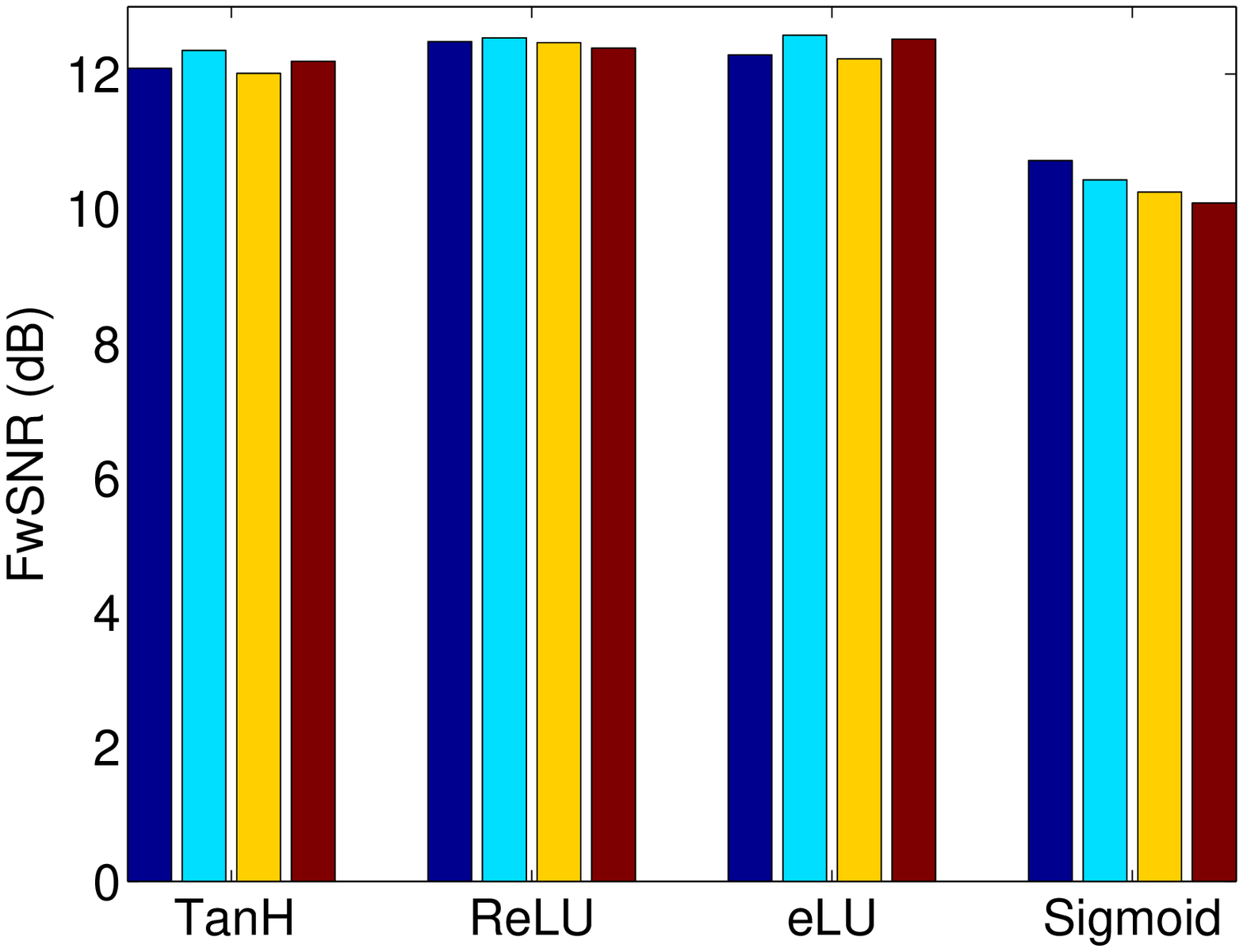}}
		\vspace{-2pt}				
		\centerline{(b)}	
	\end{minipage}
	\begin{minipage}[b]{0.48\linewidth}
		\centerline{\includegraphics[width=1.5in,height=1.1in]{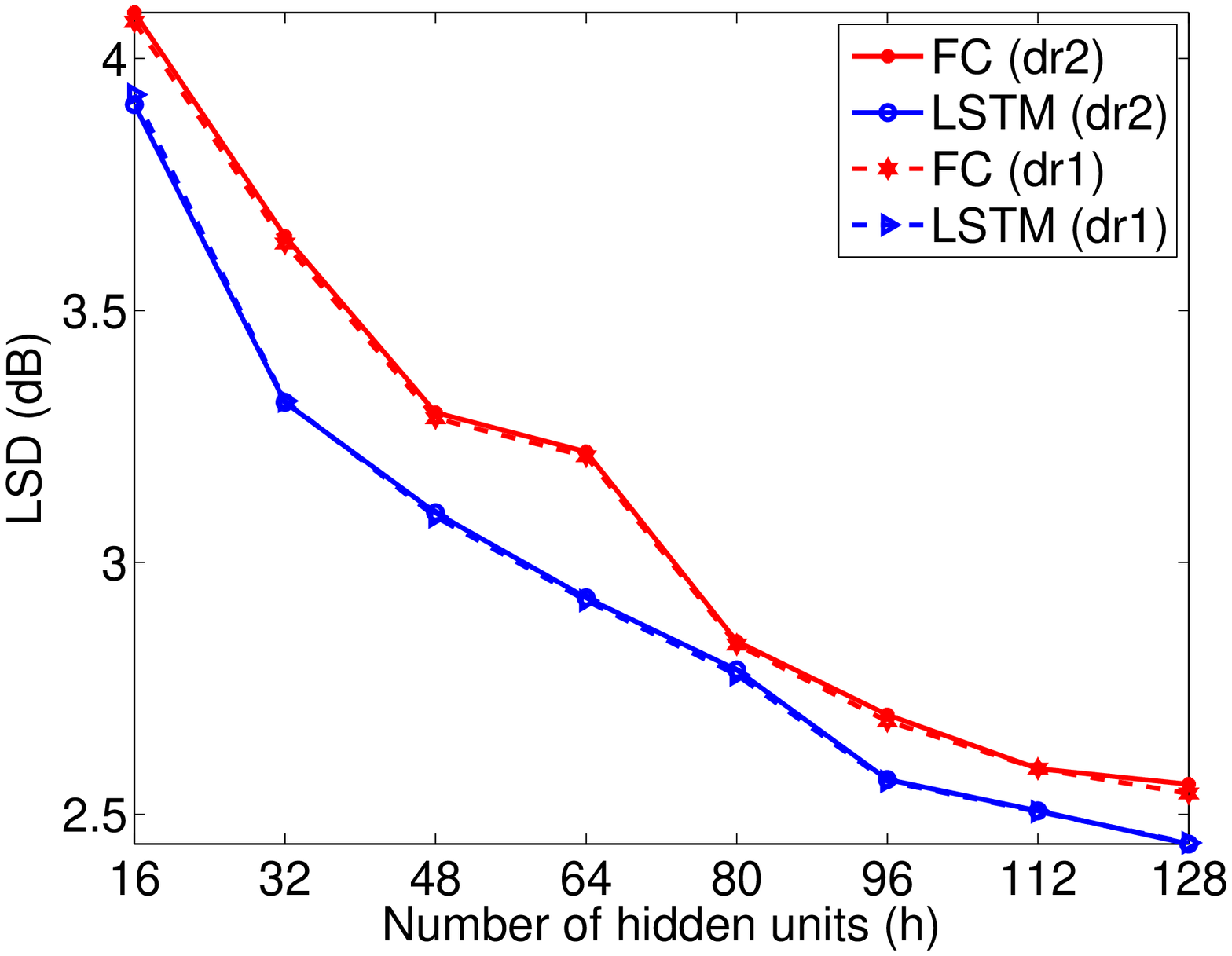}}
		\vspace{-2pt}				
		\centerline{(c)}
	\end{minipage}
	\begin{minipage}[b]{0.48\linewidth}
		\centerline{\includegraphics[width=1.5in,height=1.1in]{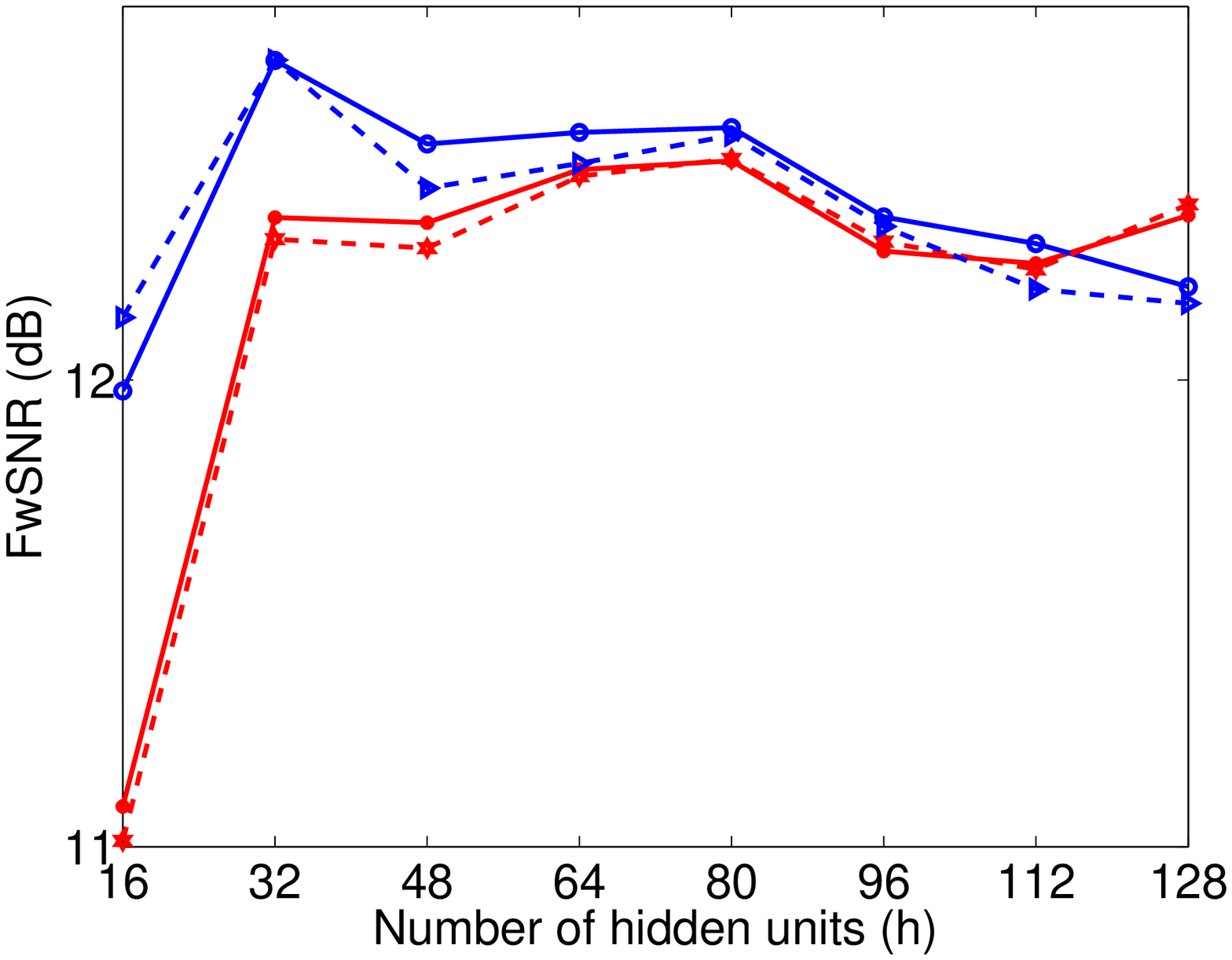}}
		\vspace{-2pt}		
		\centerline{(d)}	
	\end{minipage}		
		\vspace{-5pt}
	\caption{AE (a,c) and MCLP (b,d) performance for different activation functions (first row) and as a function of the number of units in the bottleneck layer (second row).}
	\label{fig:hiddenUnitPerformance}
\end{figure}
\par First we present the performance of the auto-encoder and then its effect on the performance of MCLP based late reverb suppression. Fig. \ref{fig:hiddenUnitPerformance}(a,b) shows the performance for different activation functions of the hidden units, for $h=64$ units in the bottle-neck layer. We see that exponential linear unit (eLU) activation performs better among the four activation functions compared; as the number of hidden units is fixed, we see smaller LSD (better AE performance) for eLU. Further, the same eLU does give a higher FwSNR for the MCLP output also (better late reverberation suppression). In contrast, the performance of Sigmoid activation is the least. The performance as a function of the number of units $(h)$ in the bottleneck layer, with eLU activation for the units is shown in \ref{fig:hiddenUnitPerformance}(c,d). We see that LSD measure does decrease with increasing the number of hidden units, which is expected since increasing hidden units increases the capacity of the network. However, increasing the bottleneck layer decreases the effectiveness of auto-encoder as a smoothing function and hence less effective as a constraint in the iterative MCLP solution. We see that the performance is better for $h$ in the range of $32-80$ hidden units and does degrade for further increase. However, for a particular $h$, LSTM is found to perform better compared to the FC architecture. 
\par From all the performance measures of Fig. \ref{fig:hiddenUnitPerformance}, we can see that the performance is similar across the two test sets `dr1' and `dr2' of TIMIT dataset, justifying the auto-encoder generalization. Among the two architectures, LSTM is clearly better for the PSD representation compared to FC architecture. We consider neural network with $h=48$ units in the bottle-neck layer and eLU activation for further evaluation. 
\begin{figure}[h]
	\centering
	\begin{minipage}[b]{0.48\linewidth}
		\centerline{\includegraphics[width=1.5in, height=1.1in]{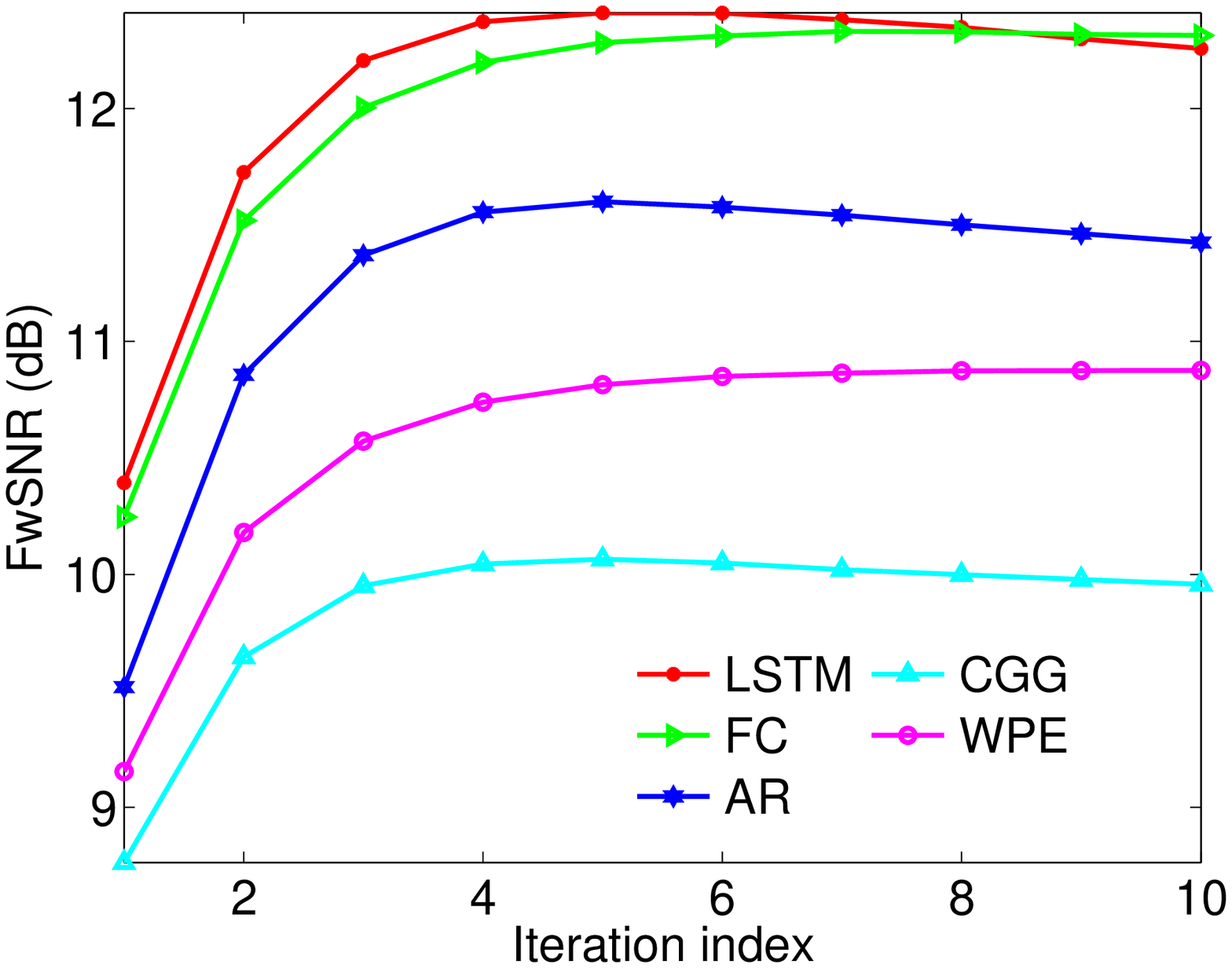}}
						\vspace{-2pt}
		\centerline{(a)}
	\end{minipage}
	\begin{minipage}[b]{0.48\linewidth}
		\centerline{\includegraphics[width=1.5in, height=1.1in]{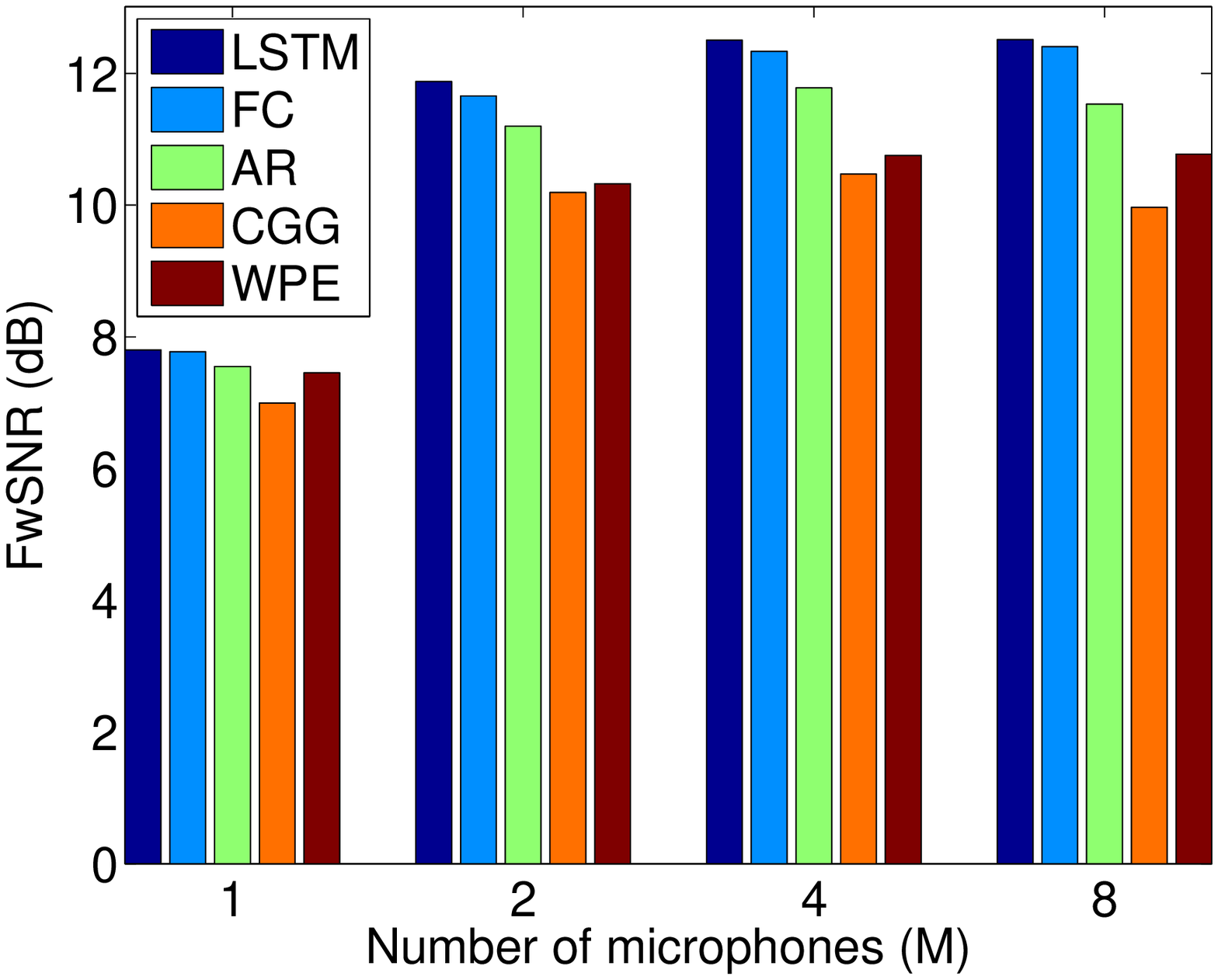}}
						\vspace{-2pt}
		\centerline{(b)}		
	\end{minipage}
	\vspace{-5pt}	
	\caption{FwSNR performance (a) as a function of the number of the iterations, and (b) for different number of microphones.}\label{fig:iterationPerformane}
\end{figure}
\par We next examine the performance as a function of the number of iterations of MCLP, shown in Fig. \ref{fig:iterationPerformane}(a). The performance increases monotonically for the first five iterations and is found to be not necessarily monotonic for all the methods. The performance in the first iteration, for which the reverberant signal is the initialization is better with the smooth PSD estimate based approaches of LSTM, FC and AR methods. This better initialization, further results in better desired signal estimation in the next iterations resulting in improved overall performance for the LSTM and FC approaches. Compared to FC architecture, LSTM is found to be better due to the temporal correlations exploited by the LSTM. We found the performance of CGG to be sensitive to the choice of the delay parameter $D$. For $D=2$ chosen in this investigation, CGG performance is poorer compared to WPE.
\par Late reverberation suppression performance for different number of microphones is shown in Fig. \ref{fig:iterationPerformane}(b). The MCLP order is chosen as $L=\{48,32,16,8\}$ for $M=\{1,2,4,8\}$ respectively, a higher order LP for smaller number of microphones $M$. Average FwSNR improves significantly for $M>1$ compared to a single microphone scenario. The performance is found to increase for $M=4$ compared to $M=2$. However, $M=8$ has similar performance compared to $M=4$. Increasing the number of microphones may also lead to degradation in the average performance, since over-parameterization may lead to over-estimation of late reflection component and hence causing signal distortion. 
\par Next, we study the performance of the proposed approach for different acoustic conditions. Table \ref{tab:perfRooms} shows the performance comparison for three different rooms and different RT60 values, and two source distances (different direct to reverberation ratio). We see that, both the original WPE method and CGG methods perform poorer compared to the MCLP-AR and MCLP-DNN priors. Performance of the AR-prior is found to be better than WPE and CGG, but poorer compared to DNN based methods. AR method estimates a smooth spectral envelope; however, for the low order prediction used traditionally the estimated envelope does not capture the harmonic information. The non-linear DNNs are better able to constrain the PSD, different from the spectral envelope and hence preserve the harmonic spectral details, resulting in better late reverb suppression. Among the two DNN schemes, LSTM is found to be better than FC network in most of the reverb examples. LSTM predicts a temporally smooth spectral prior compared to the FC architecture, resulting in better harmonic structure leading to better perceptual measures of PESQ and STOI in all acoustic conditions.
\begin{table}[h]
\caption{Late reverb suppression performance in different enclosures.}
	\vspace{-5pt}
\label{tab:perfRooms}
\begin{tabular}{|c|c|c|c|c|c|c|c|}
\hline
&         & FwSNR  & CD    & LLR   & SRMR  & PESQ  & STOI  \\ \hline
\multirow{6}{*}{ \rotatebox[origin=c]{90} {\begin{tabular}[c]{@{}c@{}}Room 1\\ 250ms\end{tabular}}} 
& Reverb	& 9.947		& 2.285		& 0.377		& 5.163		& 1.852		& 0.833\\
& LSTM		& 11.925	& 1.677		& 0.374		& 5.768		& \textbf{3.294}		& \textbf{0.927}\\
& FC		& \textbf{11.967}	& 1.694		& 0.373		& 5.778		& 3.264		& 0.922\\
& AR		& 11.512	& 1.832		& 0.431		& 5.553		& 3.110		& 0.908\\
& CGG		& 10.742	& 1.651		& 0.389		& 5.512		& 2.865		& 0.890\\
& WPE		& 11.342	& \textbf{1.637}		& \textbf{0.373}		& \textbf{5.809}		& 2.854		& 0.900\\ \hline
\multirow{6}{*}{ \rotatebox[origin=c]{90} {\begin{tabular}[c]{@{}c@{}}Room 3\\ 730ms\end{tabular}}} 
& Reverb	& 4.471		& 4.262		& 0.701		& 2.330		& 1.277		& 0.722\\
& LSTM		& \textbf{11.386}	& \textbf{1.858}		& \textbf{0.339}		& 5.669		& \textbf{2.971}		& \textbf{0.924}\\
& FC		& 11.361	& 1.896		& \textbf{0.339}		& \textbf{5.675}		& 2.901		& 0.919\\
& AR		& 11.149	& 1.872		& 0.390		& 5.439		& 2.789		& 0.908\\
& CGG		& 9.663		& 2.075		& 0.369		& 5.423		& 2.539		& 0.887\\
& WPE		& 9.552		& 2.335		& 0.400		& 5.371		& 2.285		& 0.897\\ \hline
\multirow{6}{*}{ \rotatebox[origin=c]{90} {\begin{tabular}[c]{@{}c@{}}Room 2 (Far)\\ 600ms\end{tabular}}} 
& Reverb	& 5.043		& 4.211		& 0.664		& 2.695		& 1.306		& 0.770\\
& LSTM		& \textbf{12.505}	& \textbf{1.868}		& \textbf{0.204}		& \textbf{6.164}		& \textbf{2.981}		& \textbf{0.924}\\
& FC		& 12.337	& 1.911		& 0.206		& 6.152		& 2.912		& 0.920\\
& AR		& 11.781	& 1.913		& 0.256		& 5.900		& 2.809		& 0.906\\
& CGG		& 10.472	& 2.089		& 0.229		& 5.858		& 2.560		& 0.892\\
& WPE		& 10.755	& 2.338		& 0.265		& 5.858		& 2.344		& 0.907\\ \hline
\multirow{6}{*}{ \rotatebox[origin=c]{90} {\begin{tabular}[c]{@{}c@{}}Room 2 (Near)\\ 600ms\end{tabular}}} 
& Reverb	& 10.276	& 2.613		& 0.361		& 4.618		& 2.047		& 0.956\\
& LSTM		& \textbf{14.634}	& 1.521		& 0.231		& \textbf{5.828}		& \textbf{3.752}		& 0.964\\
& FC		& 14.423	& 1.570		& 0.233		& 5.817		& 3.683		& 0.959\\
& AR		& 13.503	& 1.687		& 0.277		& 5.580		& 3.529		& 0.947\\
& CGG		& 12.918	& 1.509		& 0.227		& 5.610		& 3.262		& 0.929\\
& WPE		& 13.918	& \textbf{1.481}		& \textbf{0.222}		& 5.769		& 3.383		& \textbf{0.966}\\ \hline
\end{tabular}
\end{table}

 \vspace{-5pt}
\section{Conclusions}
\label{sec:conclusions}
 The non-linear predictive power of DNNs can be useful to improve the performance of multi-channel reverberant signal enhancement algorithms as shown in this paper. This is possible in conjunction with the iterative stochastic model based MCLP enhancement scheme. Choice of LSTM for AE-DNN and a moderate number of mic signals is found to be advantageous. The success of LSTM indicates the importance of both temporal and spectral constraint in the stochastic estimation.

\bibliographystyle{IEEEtran}
\bibliography{references}

\begin{thebibliography}{10}
\providecommand{\url}[1]{#1}
\csname url@samestyle\endcsname
\providecommand{\newblock}{\relax}
\providecommand{\bibinfo}[2]{#2}
\providecommand{\BIBentrySTDinterwordspacing}{\spaceskip=0pt\relax}
\providecommand{\BIBentryALTinterwordstretchfactor}{4}
\providecommand{\BIBentryALTinterwordspacing}{\spaceskip=\fontdimen2\font plus
\BIBentryALTinterwordstretchfactor\fontdimen3\font minus
  \fontdimen4\font\relax}
\providecommand{\BIBforeignlanguage}[2]{{%
\expandafter\ifx\csname l@#1\endcsname\relax
\typeout{** WARNING: IEEEtran.bst: No hyphenation pattern has been}%
\typeout{** loaded for the language `#1'. Using the pattern for}%
\typeout{** the default language instead.}%
\else
\language=\csname l@#1\endcsname
\fi
#2}}
\providecommand{\BIBdecl}{\relax}
\BIBdecl

\bibitem{kuttruff2016room}
H.~Kuttruff, \emph{Room acoustics}.\hskip 1em plus 0.5em minus 0.4em\relax Crc
  Press, 2016.

\bibitem{assmann2004perception}
P.~Assmann and Q.~Summerfield, ``The perception of speech under adverse
  conditions,'' in \emph{Speech processing in the auditory system}.\hskip 1em
  plus 0.5em minus 0.4em\relax Springer, 2004, pp. 231--308.

\bibitem{bradley2003importance}
J.~Bradley, H.~Sato, and M.~Picard, ``On the importance of early reflections
  for speech in rooms,'' \emph{The Journal of the Acoustical Society of
  America}, vol. 113, no.~6, pp. 3233--3244, 2003.

\bibitem{gardner1960study}
M.~B. Gardner, ``A study of talking distance and related parameters in
  hands-free telephony,'' \emph{Bell Labs Technical Journal}, vol.~39, no.~6,
  pp. 1529--1551, 1960.

\bibitem{lippmann1997speech}
R.~P. Lippmann, ``Speech recognition by machines and humans,'' \emph{Speech
  communication}, vol.~22, no.~1, pp. 1--15, 1997.

\bibitem{petrick2007harming}
R.~Petrick, K.~Lohde, M.~Wolff, and R.~Hoffmann, ``The harming part of room
  acoustics in automatic speech recognition,'' in \emph{INTERSPEECH}, 2007, pp.
  1094--1097.

\bibitem{nakatani2010speech}
T.~Nakatani, T.~Yoshioka, K.~Kinoshita, M.~Miyoshi, and B.~H. Juang, ``Speech
  dereverberation based on variance-normalized delayed linear prediction,''
  \emph{IEEE Trans. Audio, Speech, Lang. Process.}, vol.~18, no.~7, pp.
  1717--1731, Sept 2010.

\bibitem{integrated2009yoshioka}
T.~Yoshioka, T.~Nakatani, and M.~Miyoshi, ``Integrated speech enhancement
  method using noise suppression and dereverberation,'' \emph{IEEE Trans.
  Audio, Speech, Lang. Process.}, vol.~17, no.~2, pp. 231--246, Feb 2009.

\bibitem{introduction2012iwata}
Y.~Iwata and T.~Nakatani, ``Introduction of speech log-spectral priors into
  dereverberation based on itakura-saito distance minimization,'' in
  \emph{Proc. Int. Conf. Acoust., Speech, Signal Process. (ICASSP)}, March
  2012.

\bibitem{multi2015jukic}
A.~Jukic, T.~van Waterschoot, T.~Gerkmann, and S.~Doclo, ``Multi-channel linear
  prediction-based speech dereverberation with sparse priors,'' \emph{IEEE
  Trans. Audio, Speech, Lang. Process.}, vol.~23, no.~9, pp. 1509--1520, Sept
  2015.

\bibitem{jukic2015multi}
A.~Jukic, N.~Mohammadiha, T.~van Waterschoot, T.~Gerkmann, and S.~Doclo,
  ``Multi-channel linear prediction-based speech dereverberation with low-rank
  power spectrogram approximation,'' in \emph{Proc. Int. Conf. Acoust., Speech,
  Signal Process. (ICASSP)}, April 2015, pp. 96--100.

\bibitem{constrained2016jukic}
A.~Jukic, Z.~Wang, T.~van Waterschoot, T.~Gerkmann, and S.~Doclo, ``Constrained
  multi-channel linear prediction for adaptive speech dereverberation,'' in
  \emph{2016 IEEE International Workshop on Acoustic Signal Enhancement
  (IWAENC)}, Sept 2016, pp. 1--5.

\bibitem{han2015Learning}
K.~Han, Y.~Wang, D.~Wang, W.~S. Woods, I.~Merks, and T.~Zhang, ``Learning
  spectral mapping for speech dereverberation and denoising,'' \emph{IEEE
  Trans. Audio, Speech, Lang. Process.}, vol.~23, no.~6, pp. 982--992, June
  2015.

\bibitem{Williamson2017Time}
D.~S. Williamson and D.~Wang, ``Time-frequency masking in the complex domain
  for speech dereverberation and denoising,'' \emph{IEEE Trans. Audio, Speech,
  Lang. Process.}, vol.~25, no.~7, pp. 1492--1501, July 2017.

\bibitem{Kinoshita2017Neural}
\BIBentryALTinterwordspacing
K.~Kinoshita, M.~Delcroix, H.~Kwon, T.~Mori, and T.~Nakatani, ``Neural
  network-based spectrum estimation for online wpe dereverberation,'' in
  \emph{Proc. Interspeech 2017}, 2017, pp. 384--388. [Online]. Available:
  \url{http://dx.doi.org/10.21437/Interspeech.2017-733}
\BIBentrySTDinterwordspacing

\bibitem{zeiler2012adadelta}
M.~D. Zeiler, ``Adadelta: an adaptive learning rate method,'' \emph{arXiv
  preprint arXiv:1212.5701}, 2012.

\bibitem{chollet2015keras}
F.~Chollet \emph{et~al.}, ``Keras,'' \url{https://github.com/fchollet/keras},
  2015.

\bibitem{Kinoshita2016Summary}
K.~Kinoshita, M.~Delcroix, S.~Gannot \emph{et~al.}, ``A summary of the reverb
  challenge: state-of-the-art and remaining challenges in reverberant speech
  processing research,'' \emph{EURASIP Journ. on Adv. in Sig. Process.}, vol.
  2016, no.~1, p.~7, Jan 2016.

\bibitem{falk2010NonIntrusice}
T.~H. Falk, C.~Zheng, and W.~Y. Chan, ``A non-intrusive quality and
  intelligibility measure of reverberant and dereverberated speech,''
  \emph{IEEE Trans. Audio, Speech, Lang. Process.}, vol.~18, no.~7, pp.
  1766--1774, Sept 2010.

\bibitem{Rix2001Perceptual}
A.~W. Rix, J.~G. Beerends, M.~P. Hollier, and A.~P. Hekstra, ``Perceptual
  evaluation of speech quality (pesq)-a new method for speech quality
  assessment of telephone networks and codecs,'' in \emph{Proc. Int. Conf.
  Acoust., Speech, Signal Process. (ICASSP)}, vol.~2, 2001, pp. 749--752.

\bibitem{taal2010short}
C.~H. Taal, R.~C. Hendriks, R.~Heusdens, and J.~Jensen, ``A short-time
  objective intelligibility measure for time-frequency weighted noisy speech,''
  in \emph{Proc. Int. Conf. Acoust., Speech, Signal Process. (ICASSP)}, March
  2010, pp. 4214--4217.

\bibitem{SoendergaardMajdak2013}
P.~Søndergaard and P.~Majdak, ``The auditory modeling toolbox,'' in \emph{The
  Technology of Binaural Listening}, J.~Blauert, Ed.\hskip 1em plus 0.5em minus
  0.4em\relax Berlin, Heidelberg: Springer, 2013, pp. 33--56.

\end{thebibliography}

\end{document}